\documentclass[pdflatex,sn-mathphys-num]{sn-jnl}

\usepackage{graphicx}%
\usepackage{multirow}%
\usepackage{amsmath,amssymb,amsfonts}%
\usepackage{amsthm}%
\usepackage{mathrsfs}%
\usepackage[title]{appendix}%
\usepackage{xcolor}%
\usepackage{textcomp}%
\usepackage{hyperref}
\usepackage{manyfoot}%
\usepackage{booktabs}%
\usepackage{algorithm}
\usepackage{algpseudocode}
\usepackage{algorithm}%
\usepackage{algorithmicx}%
\usepackage{algpseudocode}%
\usepackage{listings}%
\usepackage{braket}

\theoremstyle{thmstyleone}%
\theoremstyle{thmstyletwo}%

\theoremstyle{thmstylethree}%

\begin{document}

\title[]{Hybrid Quantum Walks Dynamics and Construction of  a Quantum Cryptographic Primitive }


\author*[1]{\fnm{Rachana} \sur{Soni}}\email{rachanasoni007@gmail.com}

\author[2]{\fnm{Navneet} \sur{Pratap Singh}}\email{navneet.singh@bennett.edu.in}

\author[2]{\fnm{Neelam} \sur{Choudhary}}\email{neelam.coudhary@bennett.edu.in}

\affil*[1]{\orgdiv{School of Computer Science Engineering and Technology}, \orgname{Bennett University}, \orgaddress{\street{TechZone 2}, \city{Greater Noida}, \postcode{201310}, \state{U.P.}, \country{India}}}

\affil[2]{\orgdiv{School of Artificial Intelligence}, \orgname{Bennett University}, \orgaddress{\street{TechZone 2}, \city{Greater Noida}, \postcode{201310}, \state{U.P.}, \country{India}}}


\abstract{In this research article, we  design a  quantum hash function model from hybrid quantum walks on finite path graph.  The hybrid evolution operator consisting of integrated framework of continuous time quantum walks and lackadaisical quantum walks as per choice of bits in binary input message, acts on initial quantum state and generate quantum hash values from probability distribution of final quantum state.  The quantum hash values generated through the proposed method is showing strong cryptographic properties in sensitivity analysis, collision analysis, statistical features analysis, birthday attack and uniform analysis. Our proposed framework is showing successful results utilizing mathematically well defined structure of hybrid quantum walks based on quantum mechanics phenomena.}




\maketitle

\section{Introduction}

\hspace{3mm} Quantum walks are the quantum analogs of classical random walks. Initially conceptualized  by Farhi and Gutmann \cite{farhi1998quantum}. Quantum walks have become useful algorithm in quantum computing and quantum information processing. The two main version - continuous-time and discrete-time formulations can be read in ~\cite{Ambainis2001,childs2004spatial}. Quantum state transfer phenomena such as perfect and pretty good state transfer~\cite{childs2009universal} \cite{kendon2011perfect},\cite{fan2013pretty}, quantum transport~\cite{mulken2011continuous,mohseni2008environment}, and quantum fractional revival~\cite{chan2019quantum} are  helpful to researchers to find condition of transferring the quantum  states in the network with highest fidelity. Quantum walks' ability to spread information over complex networks has made them eligible candidates for quantum algorithm design~\cite{childs2009universal}, simulation~\cite{kendon2007decoherence}, and more recently, for secure cryptographic primitive designs.\\
\indent One such application is the construction of quantum hash functions (QHFs), which is proven to construct sensitive, unpredictable, and compact binary encodings of input messages. Quantum hash schemes have been proposed based on various models including discrete-time quantum walks (DTQW)~\cite{Li2013,Yang2018,Yang2019,Li2018}, alternate coin dynamics~\cite{Hou2023,Zhou2021}, Johnson graphs~\cite{Cao2018}, and even boson sampling systems~\cite{Shi2022Boson}. These schemes focus on avalanche effects, diffusion, and statistical unpredictability, which are essential features of cryptographic primitives.
Recent advances have also explored the role of memory~\cite{Zhou2021}, decoherence~\cite{Yang2021}, and hybrid mechanisms~\cite{Yang2018Dense} to improve sensitivity and randomness in QHF outputs. Furthermore, controlled alternate walks~\cite{Li2018}, cycle lattices~\cite{Shi2022}, and dihedral graph structures~\cite{Dai2018} have enriched the design space of QHF architectures.\\
\indent In this work, we propose a hybrid quantum walk framework that leverages continuous-time quantum walk (CTQW) evolution~\cite{childs2004spatial,mulken2011continuous} with lackadaisical discrete-time quantum walks (LQW), where self-loops introduce controllable laziness into the walk. The binary message serves as a control sequence determining whether the CTQW or LQW operator is applied at each step. To integrate the two quantum walk evolution  operators defined over different Hilbert space dimensions, we introduce a projection-embedding mechanism that allows CTQW dynamics to align within the extended LQW state space.\\
\indent We demonstrate that the output hashvalues generated from the final quantum state exhibits high sensitivity to small changes in input, near-uniform distribution of toggled bits, and strong segment-wise randomness. Through statistical, uniformity, and collision analyses, we establish the cryptographic strength of the proposed primitive. Our results suggest that hybrid quantum walks can serve as a foundational mechanism for secure quantum symbolic encoding and hash generation. \\
\indent Organization of the artile is as follows-  We have detailed the mathematical framework used for the proposed hybrid quantum walks scheme to generate hash values in Section 2. We perform various experimental tests to check the cryptographic strength of proposed hash function model in Section 3. We conclude the results in Section 4.

\section{Proposed Scheme}
\subsection*{Hybrid Quantum Walks Scheme to Generate QHF}

Let $G = (V, E)$ be a finite path graph with $|V| = n$ vertices. The considered path graph with number of vertices  $n=15$, and bitlength $k=11$. So total hash value length will be $15 \times 11 = 165$. The values of $n$ and $k$ can be optimized.  The quantum system evolves in a composite Hilbert space $\mathcal{H}_C \otimes \mathcal{H}_P \cong \mathbb{C}^3 \otimes \mathbb{C}^n = \mathbb{C}^{3n}$, where the coin space $\mathcal{H}_C$ has dimension 3 (left, right, and self-loop), and the position space $\mathcal{H}_P$ has dimension $n$.

We define two unitary operators:

\begin{itemize}
    \item \textbf{Continuous-Time Quantum Walk (CTQW)}:
    \[
    U_{\text{CTQW}} = \exp(-i A t) \in \mathbb{C}^{n \times n}
    \]
    where $A$ is the adjacency matrix of $G$ and $t > 0$ is a fixed evolution time.
    
    \item \textbf{Lackadaisical Discrete-Time Quantum Walk (LQW)}:
    \[
    U_{\text{LQW}} = S \cdot (C_{\text{LQW}} \otimes \mathbb{I}_n) \in \mathbb{C}^{3n \times 3n}
    \]
    where $C_{\text{LQW}}$ is a 3-dimensional Fourier coin with laziness parameter $l$, and $S$ is the conditional shift operator.
\end{itemize}

The initial state is prepared as:
\[
\ket{\psi_0} = \ket{c_0} \otimes \ket{v_0} \in \mathbb{C}^{3n}
\]

Given a binary input sequence $x = (x_1, x_2, \dots, x_k) \in \{0,1\}^k$, we define a hybrid unitary evolution:
\[
U_i =
\begin{cases}
P_{\text{embed}} \cdot U_{\text{CTQW}} \cdot P_{\text{embed}}^\dagger & \text{if } x_i = 0 \\
U_{\text{LQW}} & \text{if } x_i = 1
\end{cases}
\]

The total state after $k$ steps is:
\[
\ket{\psi_k} = U_k U_{k-1} \cdots U_1 \ket{\psi_0}
\]

\subsection*{Projection Operator for Hybrid Evolution}

To embed CTQW within the $3n$-dimensional space, we define the projection operator:
\[
P_{\text{embed}} = 
\begin{bmatrix}
\mathbf{0}_{2n \times n} \\
\mathbb{I}_{n \times n}
\end{bmatrix}
\in \mathbb{C}^{3n \times n}
\]

 To apply CTQW, we:
\begin{enumerate}
    \item Extract: $\ket{\phi} = P_{\text{embed}}^\dagger \ket{\psi} \in \mathbb{C}^{n}$
    \item Evolve: $\ket{\phi'} = U_{\text{CTQW}} \ket{\phi}$
    \item Re-embed: $\ket{\psi'} = P_{\text{embed}} \ket{\phi'} \in \mathbb{C}^{3n}$
\end{enumerate}

Thus, for $x_i = 0$:
\[
\ket{\psi_i} = P_{\text{embed}} \cdot U_{\text{CTQW}} \cdot P_{\text{embed}}^\dagger \cdot \ket{\psi_{i-1}}
\]

\subsection*{Bitstring Encoding and Final Output}

After $k$ steps, we measure the final state $\ket{\psi_k}$ in the position basis:
\[
P(v) = \sum_{c} |\braket{c, v | \psi_k}|^2
\]

Each $P(v) \in [0,1]$ is scaled and encoded into a $k$-bit binary string:
\[
\text{segment}_v = \text{bin}\left( \min\left( \left\lfloor P(v) \cdot 2 \cdot 10^4 \right\rfloor, 2^k - 1 \right) \right)
\]

The final output bitstring is:
\[
\text{QHF}(x) = \bigoplus_{v=0}^{n-1} \text{segment}_v \in \{0,1\}^{k \cdot n}
\]

This bitstring encoding defines the quantum cryptographic primitive derived from hybrid quantum walk evolution.

 \section{Performance analysis} \label{sec:sec3}
 We perform various security and strength tests for created hash values from the proposed model of the quantum hash function. 
 \subsection{Collision Analysis of the Hybrid Quantum Hash Function} \label{subsec:3.4}

We say that collision happened when hash values of two different messages are identical. Proving collision resistance in theoretical framework can be quite challenging. So we test the following experiment to evaluate collision rate.
\begin{enumerate}
    \item We choose some random  binary input   and generate the  outcome via proposed QHF scheme and represent it in ASCII format.

    \item We make modification in input data by flipping a bit randomly and generate new outcome in ASCII format.

    \item Now differentiate the both hash values in terms of identical characters  at the same position.
    \item Repeat the process $N$ times. 
\end{enumerate}

The omega count(the number of identical ASCII characters at the exact position) is given by:

\begin{equation*}
    \omega = \sum_{j=1}^N \delta \left( \tau (\xi_j ) - \tau (\xi_j') \right) 
\end{equation*}
where $\xi_j$ and $\xi_j'$ denote the $jth$ elements of the hash value of original input and  modified input, in ASCII format, respectively. $\tau (\xi_j )$ and  $\tau (\xi_j') $ denote the corresponding values in decimal format. We executed the above test several times and computed frequency for  $\omega =0$. Frequency($W_N$) for $\omega =0$ and accordingly calculated collision rates are given in the Table 1. Comparative results with other quantum hash function models are given in the Table 2. The experimental results demonstrating success and strength of proposed scheme.

\begin{table}[h]
\label{tab:omega}
\centering
\caption{$\omega$ count and collision rate table} \vspace{2mm}
\begin{tabular}{|c|c|c|c|c|}
\hline
\textbf{$\omega$} & \textbf{$W_N (N=10000)$} & \textbf{$W_N (N=20000)$} & \textbf{$W_N (N=30000)$} & \textbf{$W_N (N=40000)$} \\ \hline
0    &     9928   &  19863   &  29800 & 39719 \\ \hline

\hline
\textbf{Collision Rate}      & 0.72 \%     & 0.69  \%    &  0.67\%  &   0.71 \\ \hline
\end{tabular}

\label{tab:example}
\end{table} 
Average collision rate appears \textbf{$0.70\%$} which shows comparatively better results.
\begin{table}[h!]
\centering
\caption{Comparison of collision rate results with other QHF models.}
\label{tab:collision_resistance}
\begin{tabular}{|c|c|c|c|c|c|}
\hline
\textbf{References}            & \textbf{Ref. \cite{Yang2016}} & \textbf{Ref. \cite{Li2018}} & \textbf{Ref. \cite{Yang2018Dense}} & \textbf{Ref.\cite{Cao2018}} & \textbf{Our QHF scheme} \\   \hline \hline
Collision rate (\%)                      & 6.33                       & 9.32                       & Average 1.16               & 1.95        & Average 0.70          \\ \hline
\end{tabular}
\end{table}



 

\subsection{Statistical Analysis of the Hybrid Quantum Hash Function} \label{subsec:3.2}

\indent To evaluate the robustness and predictability of the proposed quantum hash function, we perform a statistical analysis by simulating the hash outputs over $N$ independent trials. In each trial, we give input of a random binary message  and compute its corresponding quantum hash output. We then modify a randomly chosen bit in the input message to obtain a modified message, and compute new hash values. The following statistical metrics are then recorded for each trial:

\begin{itemize}

    \item \textbf{Average Bit Difference ($\bar{B}$):} The Hamming distance $B_i = d_H(H(x), H(x'))$ is computed for each trial, and its mean over all trials gives the expected number of changed bits:($H(x)$ denotes the generated hash value for original message and $H(x'))$ is hash values for modified message) 
    \[
    \bar{\mathcal{B}} = \frac{1}{N} \sum_{i=1}^{N} B_i.
    \]
    
    \item \textbf{Bit Change Rate ($\mathcal{P}$):} The normalized percentage of changed bits in the output hash is given by:
    \[
    \mathcal{P} = \left( \frac{\bar{\mathcal{B}}}{n \cdot k} \right) \times 100,
    \]
    where $n \cdot k$ is the total output hash length in bits.
    
    \item \textbf{Variance and Standard Deviation:} We compute the standard deviation of $B_i$ (denoted $\Delta \mathcal{B}$), as well as the standard deviation of the normalized change probability, $\Delta \mathcal{P}$, to understand the fluctuation behavior of the system.

\end{itemize}

This analysis collectively examines the avalanche property and segment-wise locality of the proposed quantum hash function. A desirable hash function should yield close to 50\% of the output bits flipped, low $\Delta B$, and $\Delta \mathcal{P}$ indicating strong sensitivity and diffusion. As per the result outcome in Table 3, toggled bits counter is approximately 50 \% demonstrating that the proposed hash function generating sensitive, secure and efficient hash values.

\begin{table}[h!]
    \centering
    \caption{Statistical Analysis Test}
    \label{tab:table}
    \begin{tabular}{|c|c|c|c|c|c|}
        \hline
        & $ m =10000$ & $m=20000$ & $m=30000$ & m=40000 & \textbf{Mean} \\ \hline
        $\bar{\mathcal{B}}$ & 84.94 & 84.92&  84.84 & 84.92 &84.90\\ \hline
        $\mathcal{P}$ (\%)  & 51.48 \%  & 51.47 \% & 51.42 \% & 51.47 \% & 51.46 \%\\\hline
         $\Delta \mathcal{B} $ & 7.59  & 7.51   &  7.47  & 7.52& 7.52 \\\hline
         $\Delta \mathcal{P}$ & 4.60\%   & 4.55\%  &  4.53 \% & 4.56 \% & 4.56\\ 
        \hline
    \end{tabular}
\end{table}

     

\subsection{Sensitivity Analysis of the Hybrid Quantum Hash Function}

The sensitivity analysis shows the sensitivity of the constructed hash function for small changes in the input binary message. We consider a random binary input message data then we make some modifications in it to see the difference in outcome of hash value. 
\begin{enumerate}
    \item $C_1$ We select a random binary message and generate the hash havles.
    \item $C_2$ Insert $'0'$ bit at first position.
    \item $C_3$ flip the $5^{th} $ bit.
    \item $C_4$ Delete the $3^{th}$ bit.
  
\end{enumerate}

We constructed the following hash value in hexadecimal grouped format from the proposed hash function for the above given conditions.

 \begin{enumerate}[h!]
\item C1 (Original): ['0C02', '2184', 'BF06', 'BF35', 'D9A5', '14A4', 'FB07', '59F6', 'FD5F', 'ACA1', '241']
\item C2 (Insert bit '0' at start): ['0F4C', 'BDD1', '1398', '4537', '8C9F', '1E40', 'E672', 'CA14', '0B6D', '0C0B', '011']
\item C3 (Flip 5th): ['C6DB', 'F91E', '7079', 'E5F9', '610B', 'A7A2', '653D', 'A247', '63EC', '4A99', 'BF']
\item C4 (Delete 3rd): ['0461', '086C', '8F34', '7B6B', '6D12', 'FE89', '1DE0', 'B42D', 'A3DF', 'EECE', '4B']
\end{enumerate}
\begin{figure}[H]
    \begin{center}
    \includegraphics[width=13cm,height=6cm]{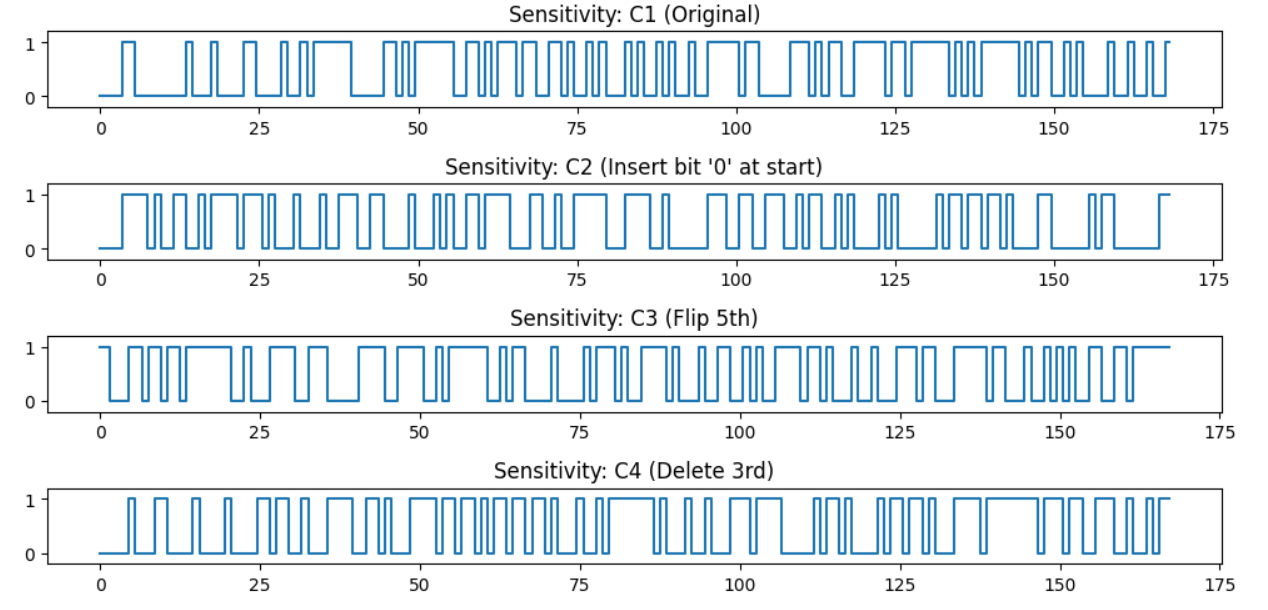}
    \caption{Sensitivity Analysis}
    \label{fig:enter-label}
    \end{center}
\end{figure}
It can be seen in the Fig. 1 that hashing scheme is sensitive towards even small change in binary input. 

\subsection{Uniform Distribution Analysis of the Hybrid Quantum Hash Function} \label{subsec:3.3}





Figure 2 illustrates the uniform distribution analysis of the proposed hybrid quantum hash function. In this experiment, we perform $40,000$ independent trials where a single-bit flip is introduced into a random binary input message, and the corresponding changes in the hash output are recorded. The histogram shows the number of times each output bit flipped across all trials.\\
\indent The resulting plot demonstrates an approximately uniform distribution of toggled bits, indicating that the proposed hash function exhibits a strong \emph{avalanche effect}. That is, each output bit is equally sensitive to small input changes, and no bit position dominates or resists change disproportionately. Such behavior is highly desirable in cryptographic applications, as it reflects good randomness, diffusion, and resistance to structural biases.

\begin{figure}[H]
    \centering
    \includegraphics[width=13cm,height=4cm]{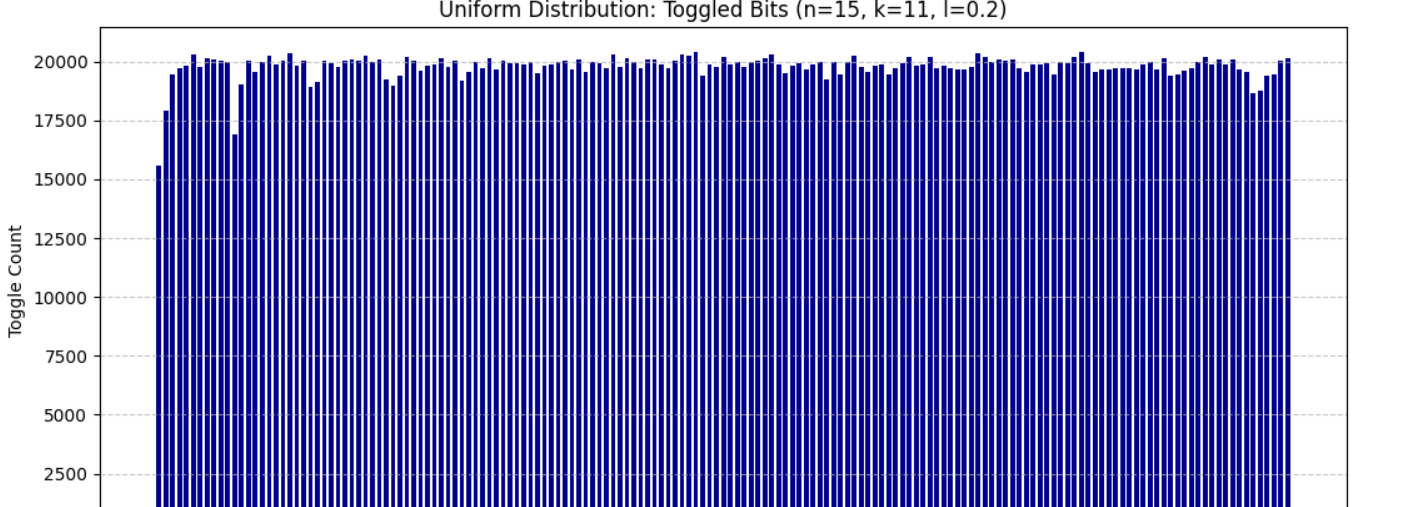}
    \caption{Uniform Distribution Analysis Test for the proposed QHF model}
    \label{fig:UniDistTest}
\end{figure}

 

 \subsection{Birthday attack} \label{subsec:3.5}

Computational complexity of Birthday Attack defines the number of trials it takes to find two input (distinct) messages map to same hash value outcome with probability of  $50 \%$. If the number of vertices in the selected graph is $n$ and bitlength is $k$ then the generated hash values length in total is $l = n \times k$. In the experiment, we have taken $n=15$ and $k=11$ so here the computational complexity of Birthday Attack is $O(2^{15 \times 11 /2})$ which is huge enough be cracked in the given time frame. Moreover, we can optimize these variables according to requirements.

\section{Conclusion}

\indent In this research study, we presented a creative mathematical framework for generating a quantum cryptographic function utilizing a hybrid quantum walks method that switches continuous-time quantum walks (CTQW) and lackadaisical quantum walks (LQW) alternatively as per the choice of bits. \\
\indent To address the dimensional differences between LQW  and CTQW, we incorporated a projection-based formalism so that evolution operators of both the walks can be matched to smoothly apply on initial quantum state. The final quantum state was projected onto the positional basis states, and the resulting probability distributions were scaled into binary segments to form symbolic output hash values.\\
\indent We conducted multiple performance analysis tests such as uniform distribution analysis to check the uniformity, sensitivity to input modifications, and segmental stability through statistical feature tests, collision test and birthday attack. These experiments exhibited strong cryptographic properties such as diffusion, high bit unpredictability, and statistical balance—key characteristics for secure hashing, showing the success of the proposed method.\\
\indent Our findings not only constructed  a novel connection between hybrid quantum walks and symbolic binary encoding but also contribute to the theoretical modeling of controlled unitary evolutions on graph-structured quantum systems. As a future work, we would like to emphasize on other underlying graph structure to explore the new possibility of quantum cryptographic primitive designs.

\section{Authors contribution} 

Rachana Soni conceptualized the study and wrote the main draft.  Navneet Pratap Singh  supervised the manuscript and Neelam Choudhary reviewed the manuscript. 
\section{Data availability statement}

This research article doesn't involve any outside dataset. All the findings are generated from the proposed research work.

\section{Conflict of interest statement}
The authors declare there is no conflict of interest. 
\section{Funding statement}
This research received no specific grant from any funding agency in the public, commercial, or not-for-profit sectors.


\bibliography{main}

\end{document}